\documentclass[12pt]{article}
\usepackage{paper2e}
\usepackage{mydefs2e}
\usepackage{epsf}

\def\caption#1{{\centerline{\vbox{\baselineskip=14pt
         \vskip.15in\hsize=5.5in\noindent{#1}\vskip.1in }}}}

\begin{document}
\begin{titlepage}

\preprint{UMD-PP-00-040, UW-PT/99-26}

\begin{center}
\Large\bf
Gaugino Mediated Supersymmetry Breaking
\end{center}

\author{Z. Chacko,$^*$\ \
Markus A. Luty,$^\dagger$\ \
Ann E. Nelson,$^*$\ \
Eduardo Pont\'on$^\dagger$}

\smallskip
\address{$^*$Department of Physics, Box 351560\\
University of Washington\\
Seattle, Washington, 98195, USA}

\smallskip
\address{$^\dagger$Department of Physics\\
University of Maryland\\
College Park, Maryland 20742, USA}

\begin{abstract}
We consider supersymmetric theories where the standard-model quark and
lepton fields are localized on a `3-brane' in extra dimensions, while
the gauge and Higgs fields propagate in the bulk.
If supersymmetry is broken on another 3-brane,
supersymmetry breaking is communicated to gauge and Higgs fields
by direct higher-dimension interactions, and to quark and lepton fields
via standard-model loops.
We show that this gives rise to a realistic and predictive model for
supersymmetry breaking.
The size of the extra dimensions is required to be of order
$10$--$100$ times larger than fundamental scale (\eg~the string
scale).
The spectrum is similar to (but distinguishable from) the predictions
of `no-scale' models.
Flavor-changing neutral currents are naturally suppressed.
The $\mu$ term can be generated by the Giudice-Masiero mechanism.
The supersymmetric CP problem is naturally solved if CP
violation occurs only on the observable sector 3-brane.
These are the simplest models in the literature that
solve all supersymmetric naturalness problems.
\end{abstract}

\date{November 12, 1999}

\end{titlepage}

\section{Introduction}
Supersymmetry (SUSY) provides an attractive framework for solving
the hierarchy problem, but it introduces naturalness puzzles of its own.
Perhaps the most serious is the `SUSY flavor problem:'
why do the squark masses conserve flavor?
A natural solution is given by models of gauge-mediated SUSY breaking
\cite{gaugemed} or by `anomalous $U(1)$' models \cite{anomU1}.
In \Ref{RS}, Randall and Sundrum suggested another solution in
theories where the visible sector fields are localized on a `3-brane'
in extra dimensions and the hidden sector fields are localized on a
spatially separated `3-brane'.
(Models of this type were introduced in the context of string theory by
Ho\v rava and Witten \cite{HW}.)
\Ref{RS} pointed out that in such theories contact terms between the
visible and hidden fields are suppressed if the separation $r$
between the visible and hidden branes is sufficiently large.
The reason is simply that contact terms arising from integrating out
states with mass $M$ are suppressed by a Yukawa factor $e^{-Mr}$ if
$M \gsim r$.
Because the suppression is exponential, the separation need only be
an order of magnitude larger than the fundamental scale (\eg~the string
scale) to strongly suppress contact interactions.

If contact interactions between the hidden and visible sector fields can be
neglected, other effects become important for communicating SUSY breaking.
One possibility is the recently-discovered mechanism of anomaly-mediation
\cite{RS,GLMR}, a model-independent supergravity effect that is always
present.
(For a careful discussion of anomaly mediation in a specific
higher-dimensional model, see \Ref{LS}.)
Unfortunately, if anomaly-mediation dominates, and if the visible sector is the minimal
supersymmetric standard model (MSSM), then slepton
mass-squared terms are negative.
This problem can be avoided in extensions of the MSSM \cite{realAMSB}.
In this paper, we will explore the alternate possibility that
standard-model gauge and Higgs fields propagate in the bulk and
communicate SUSY breaking between the hidden sector and visible-sector
matter fields. 
Models with  the all the MSSM  superfields except the gauge and Higgs fields 
localized
on a 3-brane were also considered in \Ref{SS}. In those models
supersymmetry was directly broken by the
compactification boundary conditions, requiring a rather large extra
dimension (radius of order TeV$^{-1}$) in order to explain the gauge hierarchy.
Models similar to the one considered here , {\it i.e.} 
with a hidden supersymmetry breaking sector sequestered  on a
different 3-brane and standard-model gauge and Higgs fields in the bulk, were
considered in \Ref{bulkU1}. These models contained an additional
$U(1)$ gauge multiplet;
the present paper shows that this is not required to obtain a realistic
theory of SUSY breaking.

In the higher-dimensional theory, the standard-model gauge and Higgs
fields can interact only through non-renormalizable interactions.
We therefore treat the higher-dimensional theory as an effective
theory with a cutoff $M$, which may be viewed as the fundamental
scale of the theory.
Below the compactification scale $\mu_{\rm c} \sim 1/r$, the theory
matches onto a 4-dimensional effective theory.
In this theory, the couplings of the gauge and Higgs
fields are suppressed by $1/(Mr)^{D - 4}$, where $D$ is the number of
`large' spacetime dimensions.
Therefore, the size of the extra dimensions cannot be too large
in units of the fundamental scale.
However, because the suppression of contact terms is exponential, there
is a range of radii with sufficient suppression of contact terms to avoid
flavor-changing neutral currents without exceeding the strong-coupling
bounds on the couplings in the higher-dimension theory \cite{bulkU1}.

In this scenario, SUSY breaking masses for gauginos and Higgs fields
are generated by higher-dimension contact terms between the bulk
fields and the hidden sector fields, assumed to arise from a more
fundamental theory such as string theory.
In particular, the $\mu$ term can be generated by the Giudice-Masiero
mechanism \cite{GM}.
Other direct contact interactions between the hidden and visible sectors
are suppressed because of their spatial separation.
The leading contribution to SUSY breaking for visible sector fields
arises from loops of bulk gauge and Higgs fields, as illustrated in
Fig.~1.
These diagrams are ultraviolet convergent (and hence calculable) because
the spatial separation of the hidden and visible branes acts as
a physical point-splitting regulator.
In effective field theory language, the contribution from loop momenta
above the compactification scale is a (finite) matching contribution,
while the contribution from loop momenta below the compactification
scale can be obtained from the 4-dimensional effective theory.
The higher-dimensional theory therefore gives initial conditions for
the 4-dimensional renormalization group at the compactification scale
$\mu_c$:
nonzero gaugino masses and Higgs mass parameters, and loop-suppressed
soft SUSY breaking parameters for the squarks and sleptons.
This is similar to the boundary conditions of `no-scale' supergravity models
\cite{noscale}, but in the present case the boundary conditions are justified
by the geometry of the higher-dimensional theory.
Since the SUSY breaking masses for all chiral matter fields other than the
third generation squarks are dominated by the gaugino loop, we call
this scenario `gaugino mediated SUSY breaking'
($\tilde{g}$MSB).

\begin{figure}
\begin{center}
\centerline{\epsfbox{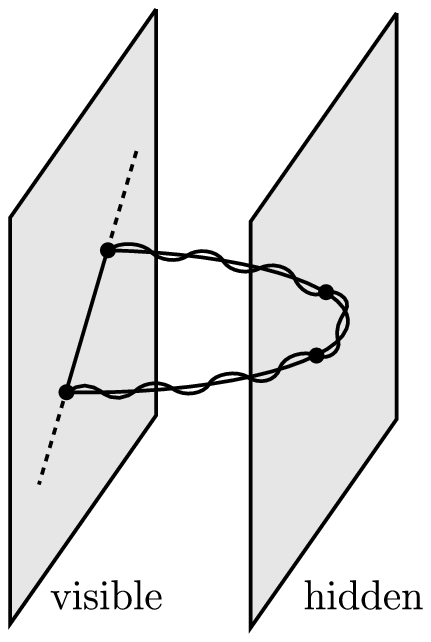}}
\smallskip
\capt{The leading diagram that contributes to SUSY-breaking
scalar masses in the models considered in this paper.
The bulk line is a gaugino propagator with two mass insertions on the
hidden brane.}
\end{center}
\end{figure}

The renormalization group has a strong effect on the SUSY breaking
parameters, and the soft masses at the weak scale are all of the same
order.
In fact, the Bino can be the lightest superpartner (LSP) in this
scenario.
The spectrum is similar to that of `no-scale' supergravity models
\cite{noscale}, with the important difference that the present scenario
allows a Fayet-Iliopoulos term for hypercharge that can have an important
effect on the slepton spectrum.
We obtain realistic spectra without excessive fine-tuning for neutralino
and slepton masses below approximately 200\GeV, suggesting that
these superpartners are relatively light in this scenario.

This paper is organized as follows.
In Section 2, we discuss the higher-dimensional theory.
We show that
the size of the extra dimensions can be large enough to suppress
FCNC's while still having gauge and Higgs couplings of order 1 at low
energies.
We also show how the SUSY CP problem can be naturally solved in
this class of models.
In Section 3, we discuss the phenomenology of this class of models.
Section 4 contains our conclusions.

\section{Bulk Gauge and Higgs Fields}
In this Section, we discuss some general features of higher-dimensional
theories with gauge and Higgs fields in the bulk and other fields localized
on `3-branes.'
We use the term `3-branes' to mean either dynamical surfaces
(\eg~topological defects or string-theory D-branes) or non-dynamical
features of the higher-dimensional spacetime (\eg~orbifold fixed
points).
All of these ingredients occur in string theory, but we will not
concern ourselves with the derivation of the model from a more
fundamental theory.
We simply write an effective field theory valid below some
scale $M$, which may be the string scale, the compactification scale
associated with additional small dimensions, or some other new physics.

We therefore consider an effective theory with $D$ spacetime
dimensions, with $3 + 1$ non-compact spacetime dimensions and $D - 4$
compact spatial dimensions with linear size of order $r$.
The $D$-dimensional effective lagrangian takes the form
\beq
\scr{L}_{D} = \scr{L}_{\rm bulk}(\Phi(x,y))
+ \sum_{j} \de^{D-4}(y - y_{j})
\scr{L}_{j}(\Phi(x, y_{j}), \phi_j(x)),
\eeq
where $j$ runs over the various branes,
$x$ are coordinates for the 4 non-compact spacetime dimensions,
$y$ are coordinates for the $D - 4$ compact spatial dimensions,
$\Phi$ is a bulk field,
and $\phi_j$ is a field localized on the $j^{\rm th}$ brane.
This effective theory can be treated using the usual techniques of
effective field theory, and parameterizes the most general interactions
of the assumed degrees of freedom below the scale $M$.%
\footnote{For an explicit supersymmetric example and calculations, 
see \Ref{MP}.}

We assume that the $D - 4$ extra spatial dimensions are compactified on
a distance of order $r \gg 1/M$.
We also assume that the distance between different branes is also of
order $r$.
This ensures that contact interactions between fields on different branes
arising from states above the cutoff are suppressed by the Yukawa factor
$e^{-M r}$.

We assume that the standard-model gauge and Higgs fields propagate in
the bulk.
Bulk gauge fields have a gauge coupling with mass dimension $4 - D$,
which is an irrelevant interaction for all $D > 4$.
When we match onto the 4-dimensional theory at the compactification
scale, the effective 4-dimensional gauge coupling is%
\footnote{We neglect the effects of gravitational curvature.}
\beq\eql{g4}
g_4^2 = \frac{g_D^2}{V_{D - 4}},
\eeq
where $g_D$ is the gauge coupling in the $D$-dimensional theory and
$V_{D - 4} \sim r^{D - 4}$ is the volume of the compact
dimensions.
If $g_D \sim 1/M^{D - 4}$, we have $g_4 \sim 1/(M r)^{D - 4} \ll 1$,
which is unacceptable.
In order to have $g_4 \sim 1$ (as observed), we require the
gauge coupling to be larger than unity in units of $M$.
However, it presumably does not make sense to take $g_D$ larger than
its strong-coupling value, defined to be the value where loop corrections
are order 1 at the scale $M$.
This follows from `\naive dimensional analysis' (NDA)
\cite{NDA,SUSYNDA}, which is known to work extremely well in
supersymmetric theories \cite{SUSYNDAwork}.
If we assume that the loop corrections are suppressed by $\ep$ at the
scale $M$, the lagrangian is \cite{bulkU1}
\beq\eql{DNDA}
\!\!\!\!
\scr{L}_{D} \sim \frac{M^D}{\ep\ell_D}
\scr{L}_{\rm bulk}(\hat\Phi / M, \partial/M)
+ \sum_j \de^{D - 4}(y - y_j) \frac{M^4}{\ep\ell_4}
\scr{L}_{j}(\hat\Phi / M, \hat{\phi}_j / M, \partial/M).
\eeq
where $\ell_D = 2^D \pi^{D/2} \Ga(D/2)$ is the geometrical loop factor
for $D$ dimensions, and all couplings in $\scr{L}_{\rm bulk}$ and
$\scr{L}_{j}$ are order 1.
Note that the fields $\hat\Phi$ and $\hat\phi$ in \Eq{DNDA} do
not have canonical kinetic terms.
The idea behind \Eq{DNDA} is that the factors multiplying
$\scr{L}_{\rm bulk}$ and $\scr{L}_{j}$ act as loop-counting
parameters (like $\hbar$ in the semiclassical expansion) that cancel
the loop factors and ensure that loop corrections are suppressed
by $\ep$.
Strong coupling corresponds to $\ep \sim 1$.

We can use \Eq{DNDA} to read off the value of the $D$-dimensional 
gauge coupling 
\beq
g_{D}^2 \sim \frac{\ep\ell_D}{M^{D - 4}}.
\eeq
We can obtain the maximum value for the size of the
extra dimension consistent with the fact that $g_4 \sim 1$
by setting $\ep \sim 1$ and using \Eq{g4}.
The results are shown in Table 1.
We see that the exponential suppression factor due to the large
size of the extra dimensions can be substantial even for many
extra dimensions \cite{bulkU1}.
Similar conclusions hold for the Higgs interactions.

\begin{table}[t] 
\centering
\begin{tabular}{c|cc|cc}
     & Torus: & & Sphere: & \\
$D$ & $M L_{\rm max}$ & $e^{-M L_{\rm max}/2}$
& $M r_{\rm max}$ & $e^{-M r_{\rm max}}$ \\
\hline
5 & 740 & $3 \times 10^{-162}$ & $118$ & $4 \times 10^{-52}$ \\
6 & 63 & $2 \times 10^{-14}$ & $18$ & $2 \times 10^{-8}$ \\
7 & 29 & $6 \times 10^{-7}$ & $11$ & $3 \times 10^{-5}$ \\
8 & 20 & $5 \times 10^{-5}$ & $8.7$ & $2 \times 10^{-4}$ \\
9 & 16 & $3 \times 10^{-4}$ & $8.0$ & $3 \times 10^{-4}$ \\
10 & 14 & $9 \times 10^{-4}$ & $7.8$ & $4 \times 10^{-4}$ \\
11 & 13 & $1 \times 10^{-3}$ & $7.8$ & $4 \times 10^{-4}$ \\
\end{tabular}
\capt{Estimates of the maximum size
and exponential suppression factor for propagation
between two branes of maximal separation.
$L_{\rm max}$ is the maximum length of a cycle of a symmetric torus,
and $r_{\rm max}$ is the maximum radius of the sphere.}
\end{table}

To see how much suppression is required, note that the dangerous
contact terms have the form (using \Eq{DNDA})
\beq
\De\scr{L}_{\rm brane} \sim \frac{e^{-M r}}{M^2}
\myint d^4\th\,
(\hat{\phi}_{\rm hid}^\dagger \hat{\phi}^{\vphantom\dagger}_{\rm hid})
(\phi_{\rm obs}^\dagger \phi^{\vphantom\dagger}_{\rm obs}),
\eeq
where the observable fields (but not the hidden fields)
have been canonically normalized.
This must be compared with the operators that give rise to the gaugino
and Higgs SUSY breaking parameters.
From \Eq{DNDA} we obtain
\beq\eql{contactops}
\bal
\!\!\!\!\!\!
\De\scr{L}_{\rm brane} &\sim \frac{\ell_D}{\ell_4} \left(
\myint d^2\th\, \frac{1}{M^{D-3}} \hat{\phi}_{\rm hid} W^\al W_\al + \hc \right)
\\
&\quad
+ \frac{\ell_D}{\ell_4} \myint d^4 \th\, \biggl\{
\frac{1}{M^{D-3}} \left( \hat{\phi}_{\rm hid}^\dagger H_u H_d + \hc \right)
\\
&\qquad\qquad\quad
+ \frac{1}{M^{D-4}} \hat{\phi}_{\rm hid}^\dagger
\hat{\phi}^{\vphantom\dagger}_{\rm hid}
\left[ H_u^\dagger H^{\vphantom\dagger}_u + H_d^\dagger H^{\vphantom\dagger}_d
+ (H_u H_d + \hc) \right]
\biggr\},
\eal
\eeq
where $W_\al$ is the gauge field strength and $H_{u,d}$ are the Higgs
fields, normalized to have canonical kinetic terms in $D$ dimensions.
(More precisely, these are $\scr{N} = 1$ superfields obtained by projecting
the bulk supermultiplets onto the branes.
For a specific example, see \Ref{MP}.)
Matching to the $D$-dimensional theory, we find
\beq\eql{initsoft}
m_{1/2}, \mu \sim \frac{\hat{F}_{\rm hid}}{M}
\frac{\ell_D / \ell_4}{M^{D - 4} V_{D - 4}},
\quad
B\mu, m_{H_u}^2, m_{H_d}^2 \sim 
\frac{\hat{F}_{\rm hid}^2}{M^2}
\frac{\ell_D / \ell_4}{M^{D - 4} V_{D - 4}}.
\eeq
Note that the $B\mu$ term and the Higgs mass-squared terms are enhanced
by a volume factor.%
\footnote{This point was missed in an earlier version of this paper.
It was pointed out in \Ref{KKS}, which appeared while this paper was
being completed.
See also \Ref{bulkU1}.}
For example,
\beq
\frac{B\mu}{m_{1/2}^2} \sim \frac{\ell_4}{\ell_D} M^{D - 4} V_{D - 4}
\sim \ep \ell_4,
\eeq
where we have imposed $g_4 \sim 1$ to obtain the last estimate.
We see that if the theory is strongly coupled at the fundamental scale,
we require a fine tuning of order $1/\ell_4 \sim 1\%$ to
obtain all SUSY breaking parameters of the same size \cite{bulkU1}.
However, for a small number of extra dimensions, the fundamental
theory need not be strongly coupled at the fundamental scale,
and we can naturally obtain all SUSY breaking terms close to the
same size.
For example for $D = 5$ compactified on a circle with circumference
$L \sim 20/M$, the exponential suppression is
$e^{-10} \sim 5 \times 10^{-5}$ and $B\mu / m_{1/2}^2 \sim 4$.
As the number of extra dimensions increases,
the strong coupling estimate is approached rapidly.
See Table 2.

\begin{table}[t] 
\centering
\begin{tabular}{c|c|c}
 & Torus: & Sphere: \\
$D$ & $B\mu / m_{1/2}^2$ & $B\mu / m_{1/2}^2$ \\
\hline
5 & 3.4 & 3.4 \\
6 & 10 & 16 \\
7 & 28 & 43 \\
8 & 69  & 85 \\
9 & 160  & 130 \\
\end{tabular}
\capt{Estimates of $B\mu / m_{1/2}^2$ for the
symmetric torus and the sphere.
The size of the extra dimension is chosen so that the exponential
suppression factor is of order
$e^{-8} \simeq 3 \times 10^{-4}$
(approximately the maximum suppression for large $D$).
This means that the torus has cycle length $L = 16/M$, and
the sphere has radius $r = 8/M$.}
\end{table}

The contribution to visible sector scalar masses from contact terms is
\beq
\De m_{\rm vis}^2 \sim e^{-M r} \left(
\frac{\hat{F}_{\rm hid}}{M} \right)^2.
\eeq
The values \Eq{initsoft} are the values renormalized at the compactification
scale;
we will later see that we require $\hat{F}_{\rm vis} / M \sim 200\GeV$.
Using the experimental constraints%
\footnote{For a complete discussion, see \eg~\Ref{SUSYFCNC}.}
\beq
\frac{m^2_{\tilde{d}\tilde{s}}}{m^2_{\tilde{s}}}
\lsim (6 \times 10^{-3}) \left( \frac{m_{\tilde{s}}}{1\TeV} \right),
\qquad
\Im \left(\frac{m^2_{\tilde{d}\tilde{s}}}{m^2_{\tilde{s}}} \right)
\lsim (4 \times 10^{-4}) \left( \frac{m_{\tilde{s}}}{1\TeV} \right),
\eeq
so $e^{-M r} \sim 10^{-3}$ to $10^{-4}$
is plausibly sufficient to suppress FCNC's.

We now discuss the loop effects that communicate SUSY breaking to the
visible sector fields, such as those illustrated in Fig.~1.
These are ultraviolet convergent because the separation of the
hidden and visible branes acts as a physical point-splitting regulator
for these diagrams.
Another way to see this is that there is no \emph{local}
counterterm in the $D$-dimensional theory that can cancel a possible
overall divergence.%
\footnote{Multiloop diagrams may have subdivergences, but these can
always be cancelled by counterterms localized on one of the branes.}
Given a specific $D$-dimensional theory, this diagram is therefore
calculable.
 From the point of view of 4-dimensional effective field theory,
the extra dimensions act as a cutoff of order $\mu_{\rm c} \sim 1/r$.
The effects of this cutoff can be absorbed into a counterterm for the
visible sector scalar masses and $A$ terms of order
\beq
\De m_{\rm vis}^2 \sim \frac{g_4^2}{16\pi^2} m_{1/2}^2,
\qquad
\De A_{\rm vis} \sim \frac{g_4^2}{16\pi^2} m_{1/2},
\eeq
where $m_{1/2}$ is the gaugino mass.
The precise value of the counterterms is calculable if we fully specify
the $D$-dimensional theory.
However, we will see that the RG running of the soft masses from
$\mu_{\rm c}$ to the weak scale gives large additive contributions to
the visible soft masses, and the final results are rather insensitive to
the precise value of the counterterm.
We will therefore be content with the simple estimate above.

Note that the soft terms arising from contact terms are larger than the
anomaly-mediated contributions, which give
\beq
\De m_{\la} \sim \frac{g_4^2}{16\pi^2} \frac{F_{\rm hid}}{M_4},
\qquad
\De m_{H_u}^2, \De m_{H_d}^2 \sim \left(
\frac{g_4^2}{16\pi^2} \frac{F_{\rm hid}}{M_4} \right)^2,
\eeq
where $M_4 \gsim M$ is the 4-dimensional Planck scale \cite{RS,GLMR}.
The other soft terms also get contributions larger than their
anomaly-mediated values from the RG, as discussed above.
Therefore, we can neglect the anomaly-mediated contribution in this
class of models.

The higher-dimensional origin of these theories can also solve
the `SUSY CP problem' \cite{SUSYCP}.
This problem arises from the fact that the phases in the SUSY breaking
terms must be much less than 1, otherwise they give rise to electron and
neutron electric dipole moments in conflict with experimental bounds.
This is a naturalness problem because CP is (apparently) maximally violated
in the CKM matrix, and it must be explained why it is not violated in
all terms.
In the present model, CP-violating phases can appear in $\mu$, $B$, and
$m_{1/2}$, generated from higher-dimension operators in \Eq{contactops}.
The phases in $\mu$ and $B$ can be rotated away using a combination
of $U(1)_{\rm PQ}$ and $U(1)_R$ transformations, leaving a single
phase in $m_{1/2}$.
This phase can vanish naturally in the present model if CP is violated only
by terms in the lagrangian localized on the visible brane.
This is a natural assumption because loop effects do not generate
local CP-violating terms in the bulk or the hidden brane.
This situation can arise (for example) if CP is broken spontaneously
by fields localized on the visible brane.
(In order to avoid a large neutron electric dipole moment, we must also
assume that the effects of the QCD vacuum angle are suppressed
\cite{strongCP}.)
%

There are many other aspects of the higher-dimensional theory that we
could discuss, but the basic features of the scenario
depend only on the qualitative feature that the visible and hidden
sectors are spatially separated.
A complete specification of the higher-dimensional model would have
to take into account the fact that there are more supersymmetries in
higher dimensions.
This may be broken spontaneously or explicitly (\eg~by an orbifold),
and couplings between bulk and boundary fields must be consistent with
SUSY.
An explicit example with 5 dimensions compactified on a
$S^1 / Z_2$ orbifold is easily constructed \cite{MP,bulkU1}.
Another important feature of the higher-dimensional theory is the
stabilization of the extra dimensions.
Stabilization mechanisms that are appropriate for the scenario we
are considering are discussed in \Refs{GW,LS}.
We conclude that there is no obstacle to constructing realistic effective
field theory models of the type outlined here.
The question of whether a model of this type can be derived from a more
fundamental theory such as string theory is left for future work.

\section{Phenomenology}
We now turn to the phenomenology of these models.
We have seen that the SUSY breaking parameters in
the effective 4-dimensional theory are determined at the compactification
scale $\mu_{\rm c} \sim 1/r$.
We have also seen that $\mu_{\rm c}$ is one to two orders of
magnitude below the fundamental scale $M$, which is most naturally
taken to be close to the string scale.
Therefore, we expect $\mu_{\rm c}$ to be close to the unification scale
$M_{\rm GUT} \sim 2 \times 10^{16}\GeV$.
We therefore identify $\mu_{\rm c}$ and $M_{\rm GUT}$ in making our
estimates.

We will further assume that the theory is embedded in a grand-unified
theory (GUT) at the scale $M_{\rm GUT}$, as suggested by the success of
gauge coupling unification in the MSSM.
We therefore consider the following SUSY breaking parameters renormalized
at $M_{\rm GUT}$:
\beq\bal
\hbox{Gaugino masses:}\ \ & M_1 = M_2 = M_3 = m_{1/2},
\\
\hbox{Higgs masses:}\ \ & m_{H_u}^2, m_{H_d}^2 \sim m_{1/2}^2,
\quad \mu, B \sim m_{1/2},
\\
\hbox{Squark and slepton masses:}\ \ & m^2 \sim \frac{m_{1/2}^2}{16\pi^2},
\\
A\ \hbox{terms:}\ \ & A \sim \frac{m_{1/2}}{16\pi^2}.
\eal\eeq
We have argued above that these conditions can emerge naturally in
this scenario for $D = 5$ or 6.
If we neglect the small loop-suppressed parameters, the model is defined by
the 6 parameters $m_{1/2}$, $m_{H_u}^2$, $m_{H_d}^2$, $\mu$, $B$, and $y_t$
renormalized at $M_{\rm GUT}$.
(We do not consider large $\tan\be$ solutions, so we neglect all other
Yukawa couplings.)
The value of $y_t$ at the weak scale fixes $\tan\be$ from the observed
value of the top quark.
The requirement that electroweak symmetry breaks with the correct value
of $M_Z$ and $\tan\be$ then fixes two more parameters.
We see that we are left with essentially 4 parameters.

An important issue when analyzing the spectrum at the weak scale
is the radiative corrections to the lightest neutral Higgs mass
\cite{radHiggs}.
The largest effect can be viewed as a top loop contribution to an effective
quartic term in the effective potential below the stop mass \cite{HH}.
We include an estimate of this effect by adding the term
\beq\eql{Hradcorr}
\De V_H = \left( \frac{3 y_t^4}{8\pi^2}
\ln\frac{m_{\tilde{t}}}{m_t} \right)
(H_u^\dagger H^{\vphantom\dagger}_u)^2
\eeq
to the Higgs potential.

We evolve the 1-loop RG equations from the scale
$M_{\rm GUT} = 2 \times 10^{16}\GeV$ down to the weak scale
$\mu_{\rm W} = 500\GeV$,
using $\al_{\rm GUT} = 1/(24.3)$.
We use input values of $m_{1/2}$, $m^2_{H_u}$, $m^2_{H_d}$, and $y_t$
at $M_{\rm GUT}$ and determine $\mu$ and $B$ by imposing
electroweak symmetry breaking.
The value of the top quark mass is used to fix $\tan\be$;
we use $m_t(\mu_{\rm W}) = 165\GeV$, which includes 1-loop
QCD corrections.
We minimize the Higgs potential including the term \Eq{Hradcorr}
with $m_{\tilde{t}}$ taken to be the heaviest of
the stop mass eigenstates, and $y_t$ renormalized at $\mu_{\rm W}$.
These approximations could be refined, but they will suffice to
illustrate the main features of the spectrum of this class of
models.

\begin{table}
\centering
\begin{tabular}{c|c|c|c|c}
&  & Point 1 & Point 2 & Point 3\\
\hline
inputs: & $m_{1/2}$ & 200 & 400 & 400 \\
& $m_{H_u}^2$ & $(200)^2$ & $(400)^2$ & $(400)^2$ \\
& $m_{H_d}^2$ & $(300)^2$ & $(600)^2$ & $(400)^2$ \\
& $\mu$ & 370 & 755 & 725 \\
& $B$ & 315 & 635 & 510 \\
& $y_t$ & 0.8 & 0.8 & 0.8 \\
\hline
neutralinos: & $m_{\chi^0_1}$ & 78 & 165 & 165 \\
& $m_{\chi^0_2}$ & 140 & 315 & 315 \\
& $m_{\chi^0_3}$ & 320 & 650 & 630 \\
& $m_{\chi^0_4}$ & 360 & 670 & 650 \\
\hline
charginos: & $m_{\chi^\pm_1}$ & 140 & 315 & 315 \\
& $m_{\chi^\pm_2}$ & 350 & 670 & 645 \\
\hline
Higgs: & $\tan\be$ & 2.5 & 2.5 & 2.5 \\
& $m_{h^0}$ & 90 & 100 & 100 \\
& $m_{H^0}$ & 490 & 995 & 860 \\
& $m_A$ & 490 & 1000 & 860 \\
& $m_{H^\pm}$ & 495 & 1000 & 860 \\
\hline
sleptons: & $m_{\tilde{e}_R}$ & 105 & 200 & 160 \\
& $m_{\tilde{e}_L}$ & 140 & 275 & 285 \\
& $m_{\tilde{\nu}_L}$ & 125 & 265 & 280 \\
\hline
stops: & $m_{\tilde{t}_1}$ & 350 & 685 & 690 \\
& $m_{\tilde{t}_2}$ & 470 & 875 & 875 \\
\hline
other squarks: & $m_{\tilde{u}_L}$ & 470 & 945 & 945 \\
& $m_{\tilde{u}_R}$ & 450 & 905 & 910 \\
& $m_{\tilde{d}_L}$ & 475 & 950 & 945 \\
& $m_{\tilde{d}_R}$ & 455 & 910 & 905 \\
\hline
gluino: & $M_3$ & 520 & 1000 & 1050 \\
\hline
sensitivity: & $m_{1/2}$ & 16 & 50 & 50 \\
& $\mu$ & 19 & 78 & 78 \\
\end{tabular}
\capt{Sample points in parameter space.
All masses are in GeV.
In the first two points, the LSP is mostly Bino, while in the
third it is a right-handed slepton.
The sensitivity parameter is defined in the main text.}
\end{table}

Some parameter choices that give rise to realistic spectra are given
in Table 3.
We find that the dependence on the overall scale of the initial
SUSY breaking masses is what would be expected:
the superpartners become heavier, and the amount of fine-tuning
required to achieve electroweak symmetry breaking increases (see below).
The right-handed sleptons get an important positive contribution from
a hypercharge Fayet-Iliopoulos term if $m_{H_d}^2 > m_{H_u}^2$ at
the GUT scale.
This distinguishes this model from `no-scale' models.
This is illustrated in the second and third parameter points in Table 3.
For $m_{H_d}^2 > m_{H_u}^2$, we easily obtain spectra where the
LSP is a neutralino.
The value of $y_t$ mainly influences the value of $\tan\be$, which
is important because the lightest Higgs boson is light for small
$\tan\be$.
We also find that $\tan\be \gsim 2.5$ is preferred in order to
have a sufficiently large mass for the lightest neutral Higgs.

An important feature of these results is the amount of fine-tuning
required to achieve electroweak symmetry breaking.
We define the fractional sensitivity to a parameter $c$ (a coupling
renormalized at $M_{\rm GUT}$) to be \cite{BG}
\beq
\hbox{sensitivity} = \frac{c}{v} \frac{\partial v}{\partial c},
\eeq
where $v$ is the Higgs VEV and the derivative is taken with all other
couplings at the GUT scale held fixed.
The largest sensitivity is to $m_{1/2}$ and $\mu$, and the values of
the sensitivity parameter are given in Table 3.
We see that the sensitivity increases strongly as the superpartner
masses are increased.
Note that even for parameters where the superpartner masses are close to
the experimental limits, the sensitivity parameter is large ($\gsim 20$).
However, it is argued by Anderson and Casta\~ no in \Ref{AC} that sensitivity
does not capture the idea of fine-tuning:
the theory is fine-tuned only if the physical quantities
significantly more sensitive than {\it a priori} allowed choices of
parameters.
 From this point of view, the fine-tuning of points with low superpartner
masses is much less severe, and naturalness clearly favors regions of
parameters with light superpartner masses \cite{AC}.
In particular, requiring that the naturalness parameter defined
in \Ref{AC} be less than $\sim 10$ implies that
the parameter $m_{1/2}$ should be less than $\sim 400$ GeV.

\section{Conclusions}
This model is the simplest supersymmetric theory in the literature
that generates an acceptable spectrum for the superpartners while
explaining the absence of non-standard flavor-changing processes and
electric dipole moments.
It is highly predictive, with squark and slepton masses qualitatively
similar to those of `no-scale' supergravity models.
The nonuniversality of the up- and down-type Higgs masses at the GUT
scale can distinguish this theory from `no-scale' 
supergravity---the expected difference between the up and down type Higgs
masses generates a hypercharge Fayet-Iliopoulos term which
affects the slepton mass spectrum.
The right-handed sleptons and the lightest neutralino are significantly
lighter than the other superpartners, and obtaining natural electroweak
symmetry breaking requires that these be lighter than roughly 200 GeV.

While this work was being completed, we received \Ref{KKS}, which
considers very similar ideas.

\section*{Acknowledgments}
M.A.L. and E.P. are supported by the NSF under
grant PHY-98-02551. Z.C. and A.E.N. are supported by the DOE under 
contract DE-FGO3-96-ER40956


\end{document}